# The Effectiveness of Computer Assisted Classes for English as a Second Language

**Ioana Iacob**
"Tibiscus" University of Timişoara, România

ABSTRACT. The present study aims to evaluate the efficiency of the computer assisted English classes and to emphasize the necessity of developing sound methodological strategies adjusted to the new technology. It also present the benefits of using the computer in the pre-school and elementary school classes, highlighted by a report on the comparative observation of four groups of children studying English in a computer assisted environment.

## 1. Aspects of the Computer Assisted English Language Learning

Language learners have unprecedented opportunities for developing second language literacy skills and intercultural understanding, in multimedia computer-assisted language learning environments. By using materials on CD-ROM, DVD's or even Web-based resources, the ESL class (English as a second language) becomes more dynamic, attention-grabbing, offering the students new entertaining ways of practicing their listening and responding skills. The premise of this study is starting from is the increase of students' motivation in solving the tasks they are given and therefore of the quality of their acquisitions.

Since the CALL (Computer Assisted Language Learning) programs appeared as a modern teaching alternative to the traditional teacher-centered ESL class, there has been a lot of concern regarding the efficiencies of this new method. Firstly, language teachers show anxiety to the idea that they might be eliminated from the process of teaching a foreign language. Secondly, the new virtual language laboratory requires for a sound training for language teachers to become familiar enough with the used technology and with the modern resources. Under the new circumstances teachers were expected to anticipate and fix the technical problems or limitations that might occur. Considering the generation gap – as a lot of language teachers has not been brought up with a computer as their students – the teachers who accept the CALL programs should make an extra effort to adjust their competence and knowledge to a completely new teaching environment.

There are some very important advantages regarding the use of multimedia programs in teaching a foreign language. First of all, it should be considered the fact that the combination of the learning paths leads to a greater success in the acquisition of the new language. Students are greatly involved visually, as in multimedia materials; the listening is combined with seeing which is more eye-catching. The student becomes very curious about what images are going to be shown on the monitor and this awakens a sense of anticipation, involving the student more actively in learning the content of the lesson. This learning environment is similar to the one in the real world where listening is combined with seeing which address a natural way of students' learning, especially when they are children. Usually the computer software used in the foreign language class is geared mainly towards the receptive skills such as listening and reading. Computer cannot assess properly the expressive skills (speaking and writing). The current computer technology limitations regard controlled speaking and writing practice. A rather successful application of automatic speech recognition is represented by the pronunciation training where the computer gives a feedback as to the accuracy of articulation. Still this facility is rather mechanic and it does not encourage the development of the speaking abilities. A valuable solution is offered by using the chat or videoconferencing to help the development of speaking and communication skills. Yet, these options that provide speaking practice, helping students in routinizing the acquired language structures are usually used outside the classroom, not involving the language teacher's control and assessment.

There are some other advantages having a psychological dimension that influences positively the learning process on the part of the learner. First of all, the computer assisted class moves the center of interest from the teacher to the learner. The student in front of the computer benefits of the advantages resulted from a learner-centered explorative approach. As the strong presence of the teacher is limited, the student feel less pressure on him/her regarding the idea that he/she might be continuously evaluated. The direct interaction student-computer redistributes the attention of the persons present in the classroom and this makes the student feel more comfortable, raising his/her confidence. The outcome is a better performance and knowledge. This way the shy or the less-able students feel free to exercise and try their responses to a task. They have more direct experiences and therefore they learn more. It is necessary to mention that the teacher's role is not eliminated. It still remains vital as the teacher selects and integrates the materials used via computer. He conducts the class, establishing specific goals, organizing review sessions, reinforcing what has been learned by monitoring the conversation, yet in a more discreet manner comparing to the traditional class. The teacher has also a very important subjective role in praising and encouraging the students' active participation which is crucial in maintaining the students' motivation.

As the students have more control over the computer assisted learning process, they can decide on the pace of learning which offers a solution to the problems raised by the differences between the slow or the fast learners. Each learner may find his own pace to get optimum results. They will not feel frustrated if not keeping up with their mates or they may not get bored by repeating some well-known items. Therefore, the use of computer program in learning English as a second language has also psychological benefits as the students feel more confident, their self-esteem will raise and they will enjoy their success. As a side benefit, they are involved in a decision making process, they find themselves in the situation of managing their learning process. Thus they get more responsible and more aware of the implications of the learning process.

Generally speaking, the use of computer technology makes the class more interesting. The language teacher makes less effort in maintaining the students focused at the learning task. Yet, there are some design issues directly related to the capacity of a particular program to catch attention and create motivation. One way a program can increase motivation in students is by personalizing the data involved. For example the student name may be integrated or some familiar contexts related to their own world of interest should be created. Images and subjects directly related to the student's life may be very efficient attention grabbers. The programs for children should make use of images and stories inspired by their in fashion world of fantasy such as cartoon, movies or stories heroes and characters. Also the program should offer practice activities incorporating challenges and stirring their curiosity and interest in performing the task correctly. Using a variety of multimedia components in one program has been shown to stimulate students' interest, increasing their motivation in spending more time on tasks with direct positive results for their achievements.

Providing attractive contexts for the use of language, the computer has a very important role in the success of the learning program. Yet, the most important factor in the efficiency of the computer assisted language teaching methods is the teacher's ability of controlling the equilibrium in the student-computer interaction. There is a risk that the student might be so excited with the form of the program and with fulfilling the tasks without paying much attention to the language involved. Some students may be fascinated by the images or by the story and they usually want to get more, getting further without seriously completing the tasks. In fact, they may mechanically fulfill their task without a proper listening to the illustrated vocabulary or without keeping in mind spelling or grammar rules. This risk regards the pre-school, elementary or secondary school children and less the college or adult students.

As the interest of the present study lies in preschool and elementary school English learners, it aims to highlight the advantages of using the computer technology in the ESL class, to take position against the opinions that invoke the risk of its limitations and to point out the main aspects leading to the successful application of this technology in the ESL class addressing children.

## 2. Advantages of using computer technology in the ESL children class

Children up to 12 are very dynamic and eager to prove their capacity of successfully fulfilling any assigned task. They usually enjoy learning, but not in a monotonous environment. They need their creativity and imagination to be challenged and they need a learning process more close to the way they naturally learn daily. The idea of sitting down at a desk and keeping the attention focused to a person or to a book is almost unnatural when it comes to children. A lesson in a book is usually illustrated by 2 or 4 pages with some static images which may look interesting for few minutes but they are not able to keep the child's interests longer. A child that has to sit down and listen is restrained to a passive position and he will lose quickly interest and motivation. The solution to this serious problem that may explain why a lot of children fail in having the traditional school success is using the computer technology. This way the child is exposed to a dynamic combination of sounds, music, images, combined in an attractive story that makes the child excited to participate to what is happening on the computer screen. Since children of this age love repetition as it addresses their need of anticipation, confirming their knowledge, it involves no effort on the part of the teacher to start over the same lesson whenever necessary to reinforce the acquisitions. Children will never get bored with this as they are given the opportunity to repeat words or sentences together with their favorite characters in a non-stressful environment as he does not feel assessed, but involved.

On the other hand, children will become keener on improving their pronunciation as their models are not represented by an adult (and it may be reasonable for a child not to be able to do what an adult is), but by characters which have a more subtle relation with their own world. A child is deeply motivated to pronounce correctly and to memorize as fast as possible the words and conversational structures used by characters such as Aladdin or Jasmine, Snow White or Mickey Mouse. Also characters that are not classics, yet establishing a relation to the children's world, being humorous or funny, have the power of giving the children enough motivation to use the English vocabulary in solving computer controlled tasks.

Therefore, the main advantage of using computer programs in the ESL class is that they maintain children attention and stimulate their motivation in actively participating to the class. Another important benefit is the fact that the child is exposed to an instance of language used by native speakers' with great effect on their correct pronunciation as it is well-known that the utterance of some specific sounds is formed during the childhood. If the children have the opportunity of using the English learning software at home, the benefits may be spectacular. The children will repeat the vocabulary daily, in short sessions, obtaining fast and accurate results. The traditional way of memorizing lists of words or of studying texts and performing "fill in the blanks" exercises has brought up a generation of rather passive English language users, with good or very good receptive skills, but poor expressive ones. By using multimedia programs, the children develop firstly their listening skill and then their speaking one. This is the natural way of learning any language, starting with the mother tongue.

It is worth emphasizing another side benefit. By interacting with the computer, children develop their computer use skills and also the sense of sharing in the collaborative classes where more children share a computer. As their language competence gets better, children will be encouraged to use English in their conversation. In time this rule will be easily assumed. On the other hand, the involved images have a teaching value themselves. They represent new and interesting information for children. For example at the lesson about farm animals, children are given the opportunity of observing the animals in their environment. They can hear real sounds from the farm and they can find out more about the activities at farm, by seeing real images or videos of this area of life maybe unknown to them.

## 3. Comparative observation of for groups of children differently exposed to CALL

The present study is based on a six months observation of the performance of four groups of children studying English as a second language: two groups of preschool children and two groups of elementary school children, about 8 children each. One group of the preschool children and one of the elementary school children have been participated to ESL classes assisted by computer, being also provided the same materials at home to be used at least two times a week. The other two groups have been participated to ESL classes assisted by computer, with no practice at home. The teacher has been very active in constantly revising the vocabulary and expanding its use in different short dialogues. The results are the following: all the groups have got good receptive skills as they understand what they are told. The difference lies is in the speed of their response and in its accuracy. The groups that used the software at home presented a significant decrease in their reaction time for word recognition and for giving answers at specific questions. They also show a greater vocabulary gain and their usage of words is more correct and fluent. Therefore, it is recommended that the computer software used at school to be also used at home for fast and accurate results.

## 4. Argument against the significance of limitation of the computer technology in the ESL children class

The adoption of computer assisted language learning in schools has determined some criticism as to the real efficiency of the method. Some of the highlighted limitations are: the computer cannot properly evaluate the learner's speech, it mainly addresses the receptive skills, it is costly, it cannot provide consistent feedback, and so on. These limitations are real, yet they are to be compensated by the teacher's active role. The computer technology is not meant to replace the teacher, but to assist him/her. The teacher is the one who should keep the children interest in the language task, not allowing them to invest more of their energy in watching the "form" and missing the valuable content. The language teacher should provide constant feedback; he should make sure the child is gasping the idea by asking questions on the new vocabulary, he should praise the children's success and provide new communication contexts to practice the new vocabulary.

The flexibility in creating new contexts for using the vocabulary and in helping the child to make correct use of the new words is an irreplaceable human quality. Only the teacher can provide situations in which the poly-semantics and poly-functionality of a word can be understand and exercised. For example, the verb *to take* is part of some fix structures such as *to take a shower, to take a bath, to take a picture, to take a ride* and so on. The role of the teacher is to practice these structures in different dialogues with the children. While the computer may develop especially the receptive skills, the teacher' role is to develop the expressive ones which are much more related to the human dimension of communication.

**Conclusions**

In conclusion, the real issue of the computer assisted language learning is namely the lack of sound pedagogic strategies of this learning system. A consistent methodological and pedagogical research should be carried out regarding the use of computer in language classes. In fact the methodology lags behind the software that recently has appeared on the market. Therefore, the limitations are not related to the computer software, but to the assessment of the teaching methods that should be involved. While this problem is to be approach, the reality of the measurable effectiveness of computer assisted training for children learning English as a second language cannot be denied.